# Crack Propagation in Bone on the Scale of Mineralized Collagen Fibrils: Role of Polymers with Sacrificial Bonds and Hidden Length

Wenyi Wang[1] and Ahmed Elbanna[1]

[1]Department of Civil and Environmental Engineering
 University of Illinois Urbana Champaign
 205 N. Mathews Ave
 Urbana, IL 61801, USA

 Contact corresponding author:
 Phone:+001(217)9796050
 Email: wwang72@illinois.edu

 Co-corresponding author:
 Phone:+001(217)7512117
 Email: elbanna2@illinois.edu

# Abstract

Sacrificial bonds and hidden length (SBHL) in structural molecules provide a mechanism for energy dissipation at the nanoscale. It is hypothesized that their presence leads to greater fracture toughness than what is observed in materials without such features. Here, we investigate this hypothesis using a simplified model of a mineralized collagen fibril sliding on a polymeric interface with SBHL systems. A 1D coarse-grained nonlinear spring-mass system is used to model the fibril. Rate-and-displacement constitutive equations are used to describe the mechanical properties of the polymeric system. The model quantifies how the interface toughness increases as a function of polymer density and number of sacrificial bonds. Other characteristics of the SBHL system, such as the length of hidden loops and the strength of the bonds, are found to influence the results. The model also gives insight into the variations in the mechanical behavior in response to physiological changes, such as the degree of mineralization of the collagen fibril and polymer density in the interfibrillar matrix. The model results provide constraints relevant for bio-mimetic material design and multiscale modeling of fracture in human bone.

# 1. Introduction

The human bone structure is a hierarchical composite of collagen and hydroxyapatite (HA) with several mechanisms to resist fracture at various scales [Koester et al., 2008; Nalla et al., 2004; Ritchie et al., 2009] .These size scales relate to the characteristic structural dimensions in bone, which vary from twisted peptide chains at the nanometer scale to the (secondary) osteon (haversian) structures, which are several hundred micrometers in size. The hierarchical structure at the intermediate scales includes (i) hydroxyapatite-impregnated twisted collagen fibrils at the scale of tens of nanometers; (ii) collagen fibers that are typically a micrometer in diameter and (iii) the lamellar structure of collagen fibers at several micrometer dimensions. The combination of this complex geometry and unique blending of material properties provides bone with remarkable levels of strength and toughness [Espinoza et al., 2009; Elbanna and Carlson, 2013].

In this paper, we focus on the mechanical response of a single mineralized collagen fibril sliding on a polymeric layer that includes sacrificial bonds and hidden length (SBHL) systems [Thompson et al., 2001]. The fibril utilizes the breakage of sacrificial bonds and the release of hidden length to dissipate energy while being stretched. This process introduces a microscopic mechanism for fracture resistance [Fantner et al., 2005]. Our primary focus is investigating the effect of the polymeric glue material on the basic characteristics of crack propagation such as critical crack size, stable crack growth speed and energy dissipation.

The basic structure and operation mechanism of the SBHL system is shown in Figure 1. The assembled glue molecule may include more than one polymer chain with sacrificial bonds forming within the chain itself, crosslinking the different chains and connecting the chains to the collagen fibrils. The large scale separation of the collagen fibrils is resisted by an array of parallel gel molecules as shown in Figure 1(a). As long as the bond is intact, it shields parts of the polymer length from contributing to the end-to-end distance. This corresponds to a reduction in the chain entropy (the possible number of configurations resulting in the same end-to-end distance) and a corresponding increase in the initial stiffness of the polymer chain. After the sacrificial bond is broken, the shielded loop unfolds and significant energy is dissipated in reducing the chain entropy as it straightens out.

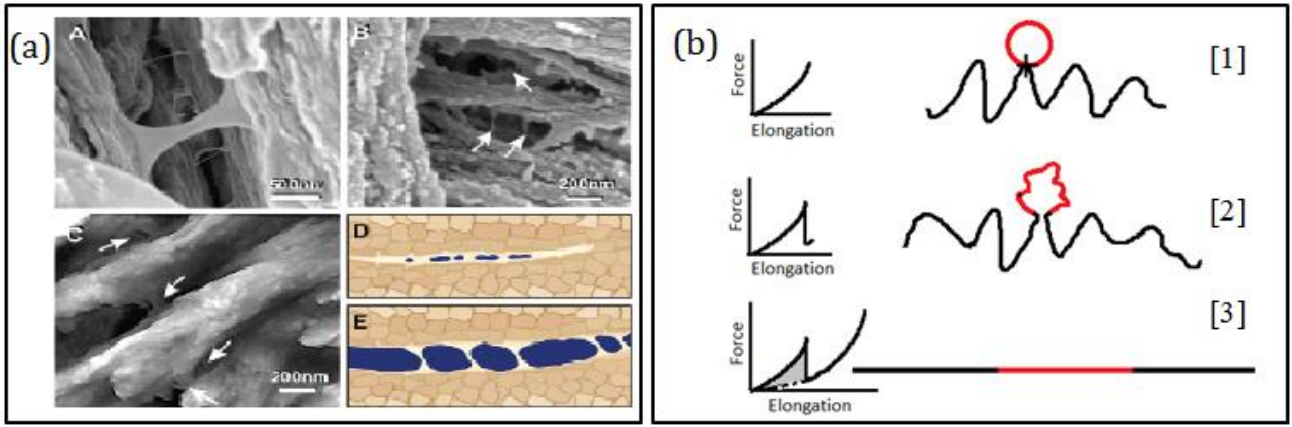

Figure 1: The structure and the basic operation principles of the SBHL system. (a) High-resolution Scanning Electron Micrographs (A and B) and AFM (C) show glue molecules resisting fracture in bone suggesting that these molecules form quasi-one-dimensional bundles. Subplots (D) and € shows two adjoining mineralized collagen fibrils at rest and during the formation of microcracking respectively.  (Reprinted with permission from G. E. Fantner et al., [2005]). (b) Force change associated with sacrificial bond breakage and hidden length release. (i) Before a sacrificial bond is broken, only the black length of the molecule contributes to the entropic configurations and to the force with which the molecule resists stretching. The red length of the molecule is hidden from the force by the sacrificial bond. (ii) When the bond breakage threshold is reached, the bond breaks and the whole length of the polymer (black plus red) contribute to its configurational entropy. This sudden increase in entropy leads to an abrupt force drop. (iii) As the polymer molecule is further stretched, the force it supports increases, until the entire molecule detaches from the substrate and ruptures. The grey area represents the extra work done in stretching a polymer with sacrificial bonds and hidden lengths, relative to a polymer of the same length but without such structural features (from Elbanna and Carlson [2013]).

Previous theoretical models describing the mechanical behavior of bone glue polymers [e.g. Fantner et al., 2006; Elbanna and Carlson, 2013; Lieou et al., 2013] have implemented the worm-like chain model [Bustamente et al., 1994] as an approximation for the AFM experimental curves. We adopt this model here as well. The more flexible the glue polymers are, the better this approximation will be. Nonetheless, further work is required to constrain the force displacement relation of single polymer molecules in the bone glue along its whole deformation history.

The existence of SBHL systems is incorporated in the worm-like-chain model by introducing a dynamical variable: the available length [Elbanna and Carlson, 2013]. This available length is the difference between the polymer contour length and the sum of the length of the hidden loops that have not been unfolded yet. The rate dependence of the SBHL system is modeled using the transition state theory [Bell, 1979; Lieou et al., 2013]. In this paper we will implement the rate and displacement model developed by Lieou et al., [2013] as the constitutive law for the polymeric layer with SBHL system.

The primary component of the human bone structure is mineralized collagen fibril. Buehler [2007] developed a model for the mineralized fibril in nascent bone in which collagen is represented by tropocollagen molecules, cross-linked by hydrogen bonds, and the mineral plates are hydroxyapatite (HA) crystals forming in the gap regions between the collagen fibrils. The stiffness of collagen fibrils depends on the mineralization percentage. With aging, bone properties degrade [e.g. Zioupos and Currey, 1998]. The mineralization percentage decreases and both the stiffness and the peak strength of the fibrils are reduced. We will also study the influence of mineralization on fracture properties of the mineralized fibril-polymer system.

We developed a coarse grained model for the mineralized collagen fibril with polymeric glue. The fibril is modeled using a one-dimensional mass-spring system. The stiffness of the springs is calculated using the fibril geometric properties and the stress strain behavior computed from Beuhler [2007]. The polymeric layer is modeled using the constitutive description of Lieou et al., [2013]. The system is integrated in the quasistatic limit which is appropriate for exploring nucleation characteristics and stable crack growth speeds. Depending on the polymer density, the system may fail by the breakage of the collagen fibril and not the detachment of the polymers. In this limit we use a fully dynamic approach to track these instabilities. This failure mode is relevant for understanding the deterioration of bone quality with age since the ability of bone cells to produce the polymeric glue decreases with age. We have also investigated the properties of the SBHL system since it has been shown previously that these molecular bonding provide a small scale energy dissipation mechanism and hence contribute to fracture toughness [Thompson et al., 2001]. There is still insufficient experimental data about the geometrical properties of these systems and the effect of this mechanism on crack resistance. Hence, we pursue in this paper a parametric study to explore their relative contribution on fracture processes.

The remainder of the paper is organized as follows: In Section II we introduce our model for numerical simulation as well as its discretization. Then, the material properties of collagen fibrils and the polymer system are discussed. In Section III, we describe the numerical method and the integration scheme. In Section IV, the results of our simulations are presented demonstrating the effect of different properties of SBHL system, polymer density and mineralization ratios. We discuss the implications of our simplified model on bone fracture in Section V.

## 2. Kinetic model

In this section, we introduce the basic elements of the coarse grained model for the mineralized collagen fibril and the polymeric layer. We consider a single fibril, idealized as 1-D array of masses and nonlinear springs, sliding on a viscoplastic polymeric layer. As the fibril is pulled, the motion is resisted by the interfacial forces provided by the polymeric system. Detachment of polymer end bonds and failure of collagen fibril are expected as limit states.

**2.1 Model setup**

A mineralized collagen fibril may be idealized as a 1-D prismatic solid bar. We are primarily interested in the longitudinal deformation of the bar as interfibrillar slip is one of the major failure modes in fibrillar arrays [Beuhler, 2007; Espinosa, 2009]. Polymer molecules are uniformly distributed along the interface. Displacement controlled loading is applied in the longitudinal direction on one end of the collagen fibril as shown in Figure 2(a). We discretize the collagen fibril into a number of identical blocks connected with nonlinear springs which capture the behavior of the mineralized fibril molecules as shown in Figure 2(b).

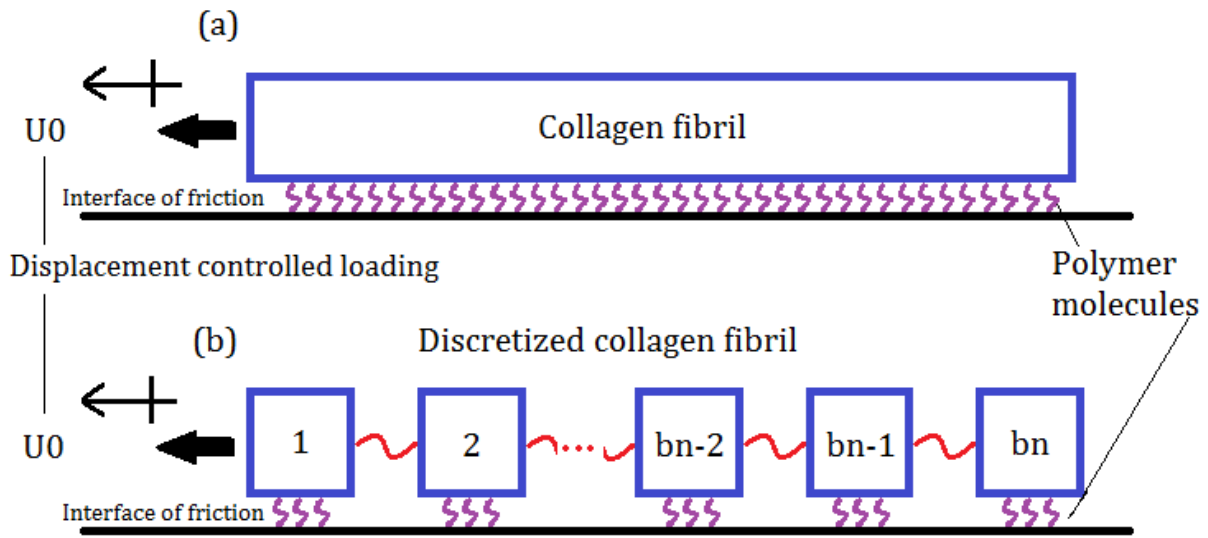

Figure 2: The mineralized fibril. (a) Schematic plot of the continuum representation (b) Schematic plot of discretized model, $b_n$ is the number of discretized fibril blocks.

The stiffness of the springs interconnecting the blocks is computed based on the geometric and material properties of the mineralized fibrils. The material model for the fibril is adapted from Beuhler [2007] (Figure 3(a)). The different stress drops represent internal slip events between the tropocollagen molecules or between collagen molecules and mineral plates. We adopt a simplified version of these curves with a linear elastic behavior up to the yield point followed by a saw tooth response in the post-yielding phase up to the point of complete failure (Figure 3(b)).

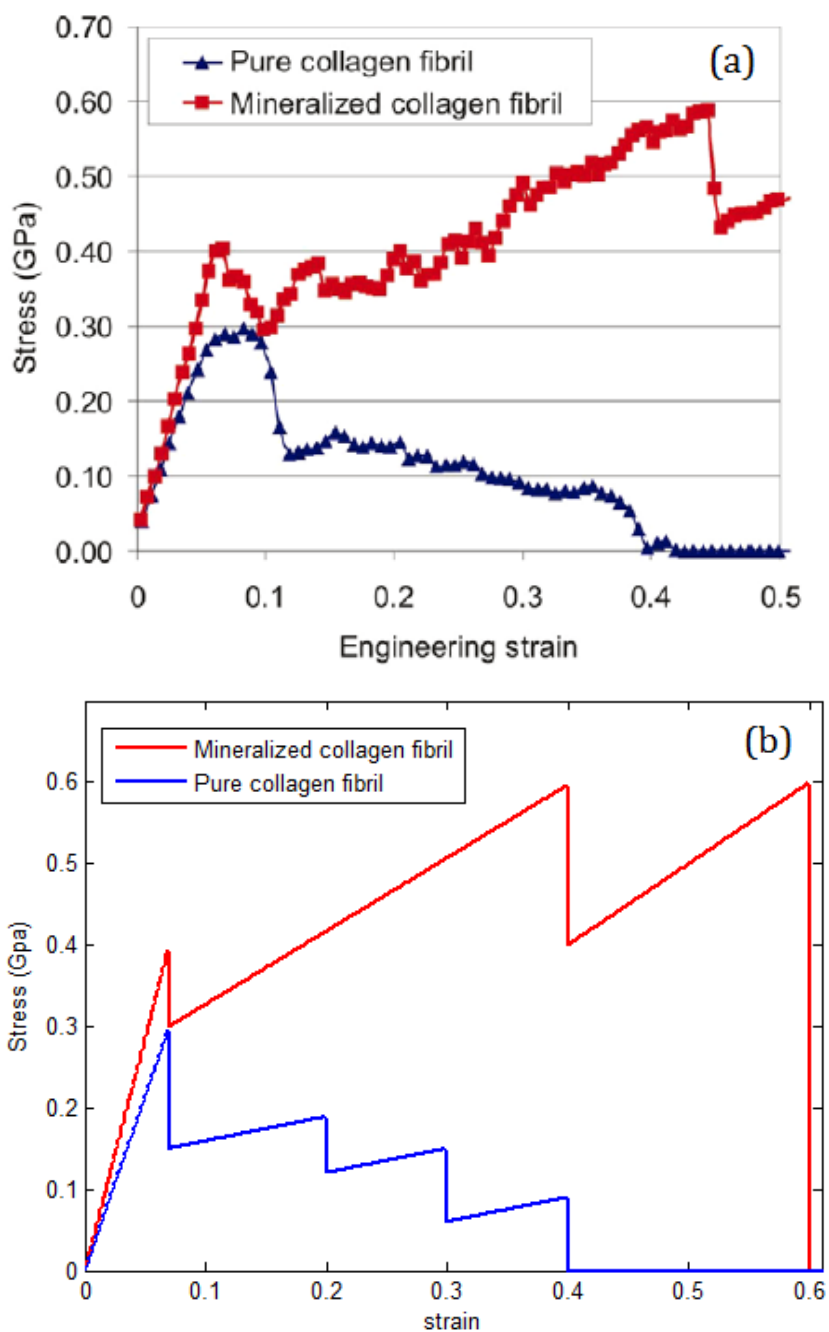

Figure 3: Constitutive behavior of collagen fibrils. (a) Stress-Strain relation of mineralized collagen fibril & pure collagen fibril [Beuhler, 2007]. Figure reproduced with permission from Beuhler [2007] (b) Simplified stress-strain relation of mineralized collagen fibril & pure collagen fibril.

For the stress-strain relation under compression, we assume that the fibril is linear elastic up to its buckling stress and has no compressive strength in the post-buckling regime. We take the buckling stress in compression to be equal to the tensile yield stress. That assumption is sufficient for our purpose, because simulations show the compression strain of collagen fibril will not exceed -0.07; the assumed value for yield strain. We discuss the implications of this assumption and possible modification in Section V.

If the fibril is unloaded in the post-yielding regime, residual inelastic strains develop and energy is dissipated. Beuhler [2007] found that the energy dissipation ratio between mineralized collagen fibril and pure collagen fibril is approximately 5. Since the full stress strain curve for the unmineralized collagen fibril is known, we use this estimate of the energy dissipation to extrapolate the stress strain curve for the mineralized case beyond what is shown in Beuhler [2007] (Figure 3(a)).

**2.2 Dynamical constitutive behavior of polymer system**

Based on the worm-like chain model, the force extension relationship for a single polymer is given by [Bustamente et al., 1994; Rubenstein, 2003; Elbanna and Carlson, 2013]:

$$F = \frac{k_B T}{b} \left[ \frac{1}{4}\left(1 - \frac{x}{L_a(x,\dot{x})}\right)^{-2} - \frac{1}{4} + \frac{x}{L_a(x,\dot{x})} \right] \qquad (1)$$

where $F$ is the polymer force, $x$ is the end-to-end distance, $\dot{x}$ is the pulling rate, $b$ is the persistence length, $k_B$ is Boltzmann constant, $T$ is the temperature and $L_a$ is the available length of polymer, which is the sum of the length of polymer parts that contributes to entropic elasticity [Elbanna and Carlson, 2013]. The available length depends implicitly on the pulling rate. To account for this rate effect, we adopt the rate and displacement model of Lieou et al. (2013). This model uses the transition state theory to construct a master equation of the bond breakage rates and thus provides a tool to track the variations in the available length. We review this approach in Appendix A1.

The polymeric system is usually composed of a large number of polymers (*Np*). We assume that the polymer molecules are parallel to one another forming quasi-one-dimensional bundles and neglect cross linking between the bundles. Each idealized polymer molecule, however, may consist of more than one single polymer strand crosslinked together as shown in Fig. 1a. In this case, the hidden length concept does not represent globular domains only but is also extended to cover parts of the polymer chains that are shielded by the cross links. Also, the persistence length used in the WLC model should be interpreted as an effective persistence length for the polymer blob to fit its mechanical behavior and not necessarily the actual persistence length of a single strand. The blob of polymer molecules deforms predominantly in one dimension and thus we approximate it by a quasi-one-dimensional bundle. Figure 1a suggests that during the separation of the fibrils, several of these bundles resist the failure with weak connection between the bundles. We thus neglect the cross linking between the fiber bundles. This approximate model of polymers with SBHL has successfully reproduced many of the features observed in previous AFM experiments [e.g. Fantner et al., 2006; Elbanna and Carlson, 2013; Lieou et al., 2013]. It is possible, nonetheless, that the actual topology of the polymeric interface is more complicated. In particular, the cross linking between the polymer molecules may lead to the development of two-dimensional network structure. We discuss this further in Section V.

Moreover, we assume that the contour lengths of polymers as well as the lengths of the hidden loops are chosen from a uniform random distribution. By coupling the worm-like chain model equations with the transition rate factors (See Appendix A1 for details) it is possible to generate force extension curves for the polymeric system at different pulling rates. This is shown in Figure 4 for both the single and multiple polymers cases. If all polymers are detached, we assume that the residual frictional resistance of the polymeric layer is negligible. The rate dependence of the residual friction will be the subject of future investigation.

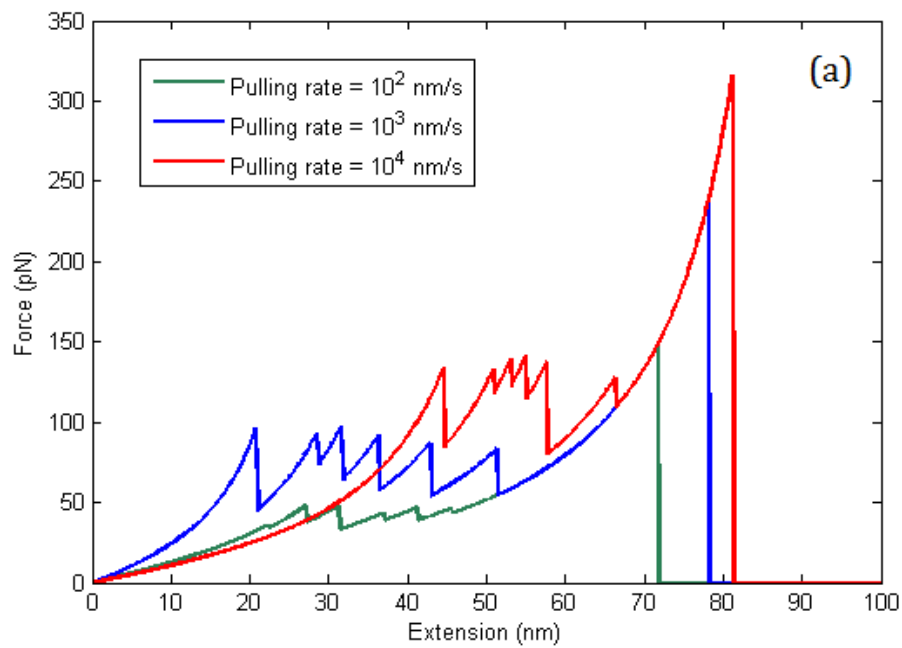

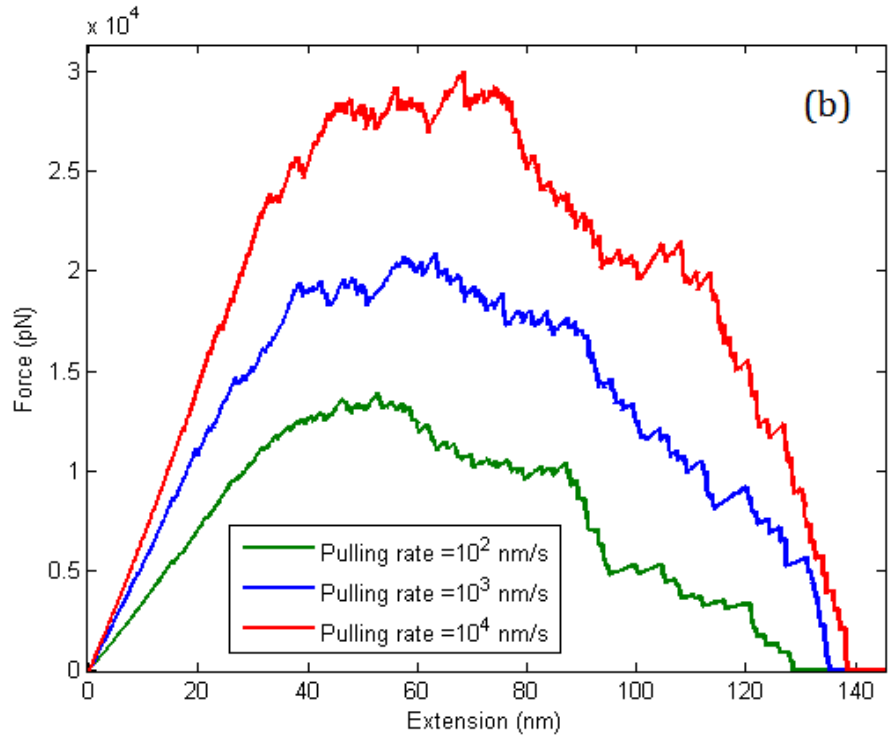

Figure 4: Constitutive response of the polymer molecules, stretched at v=10$^2$m/s (green), v=10$^3$m/s (blue) and v=10$^4$m/s (red). (a) Single polymer molecule force-extension curve. (b) Polymer system force-extension curve.

Additionally, we allow for the case of polymer retraction. That is, when the pulling rate becomes negative, the system unloads. The unloading force decreases according to the worm-like Force-Elongation model (Eqn. (1)), now applied to a group of polymers simultaneously with their available lengths the same as their values at the unloading point (i.e. we assume no bonds break during unloading). This is shown in Figure 5.

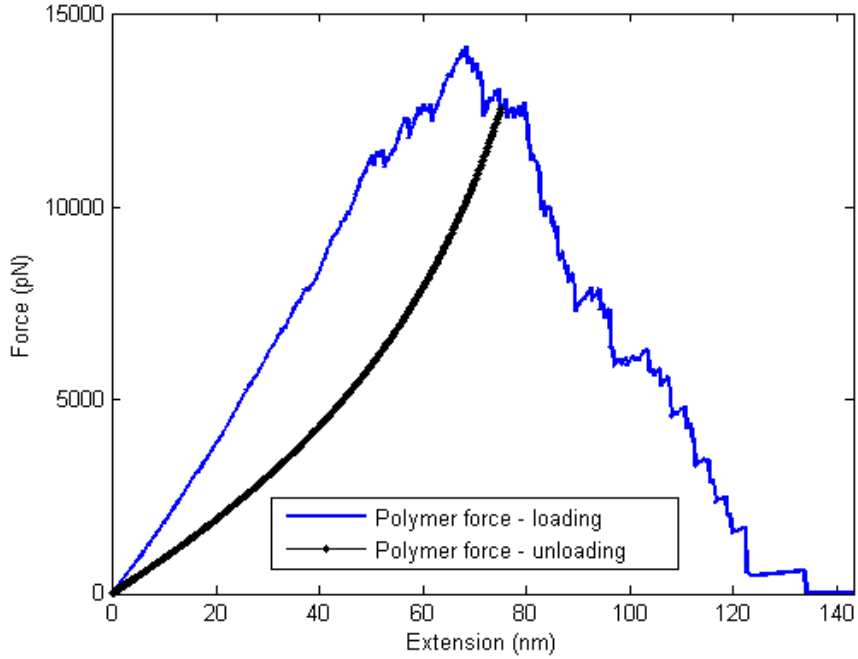

Figure 5: Polymer force under loading (Blue) and unloading (Black) in a representative numerical experiment.

**2.3 Parameters selection**

In the numerical simulations we used the following parameters for the collagen fibril model. We assume that its cross section is square with side dimension $l_s$= 100 nm. The total length of the fibril bar is $L_f$= 2000nm, and is discretized into $b_n$ = 100 identical blocks with each length = 20nm. The density of fibril is assumed to be $\rho$=1500 kg/m$^3$. We apply a constant pulling rate of $v_0$ =1 µm/s at the first block. The elastic modulus $E$ of the fibril is approximately 5.7 GPa for the fully mineralized case and 4.3 GPa for the unmineralized case. The stiffness of the spring is calculated as $k = EAb_n / L_f$. The yielding strain is $\varepsilon$=0.07 for both tension and compression.

We assume the polymer density is spatially homogeneous along the interface. We consider different values for this density $D$ varying between 5 and 25 polymers per nm. Each polymer is assumed to have the same number of sacrificial bonds (N). The effect of increasing the number of sacrificial bonds per polymer from 0 to 8 will be discussed shortly. Setting a uniform number of sacrificial bonds across the polymer enables us to investigate the effect of number of SBHL on crack propagation independent of other factors. In real systems it is possible that the number of sacrificial bonds may vary from one polymer bundle to another. This is accounted for, partially, in our simulations by allowing randomness in other system variables (e.g. contour length and length of hidden loops) as we will shortly discuss. Furthermore, previous work [e.g. Elbanna and Carlson, 2013] has shown that the maximum increase in toughness from a polymer blob is achieved with just a few sacrificial bonds (less than 10). The contour lengths of polymers ($L_c$) are chosen from a uniform random distribution with average length, minimum length and maximum length of 150 nm, 75nm and 225 nm, respectively. These values are consistent with what is observed in AFM experiments [Hansma et al., 2005]. The hidden lengths ($L_n$) are also generated randomly between 0 to $C*L_c / N$ with the only constraint that the initial available length ($L_0$) is positive, where $C$ is a positive design coefficient typically set as unity.

All the parameters are summarized in Table 1.

| Parameter | Physical meaning | Value in simulation | References |
|---|---|---|---|
| $l_s$ | Collagen fibril cross section side dimension | 100 nm | Ritchie et al., 2009 |
| $L_f$ † | Length of collagen fibril bar | 1000nm<br>2000 nm* | Ritchie et al., 2009<br>Beuhler 2007<br>*Assumed in simulation |
| $b_n$ | Number of discretized blocks in fibril bar | 100 | Assumed in simulation |
| $\rho$ | density of fibril bar | 1500 kg/m$^3$ | Kurtz, SM. Edidin AA 2006 |
| $v_0$ ‡ | Loading velocity | 1 µm/s | Assumed in simulation |
| $E$ | Elastic modulus of collagen fibril | 5.7 / 4.3 GPa (mineralized / unmineralized) | Beuhler 2007 |
| $\varepsilon$ | Yielding strain of collagen fibril | 0.07 | Beuhler 2007 |
| $D$ | Polymer density | 5 ~ 25 /nm | Varies in simulation |
| $N$ | Number of SBHL in each polymer | 0 ~ 8 | Varies in simulation |
| $L_c$ | Contour length of polymer molecule | 75 ~ 225 nm | Lieou et al 2013;<br>Hansma et al., 2005 |
| $L_n$ | Hidden length of each hidden loop | 0 ~ $C*L_c / N$ | Varies as uniform distribution in simulation |
| $L_0$ | Initial available length of polymer molecule | >0 | Determined by $L_c$ and $L_n$ |

Table 1. List of parameters used in model and simulation

†: $L_f$ has been estimated previously to be of the order of 1 micrometer [Ritchie et al., 2009]. However, we extended the bar to 2 micrometers to ensure that there is no effect for the end conditions on the crack characteristics in the stable growth regime. We got identical results for the critical crack size and stable crack growth for simulations with a bar of length 1 micron (not shown here). The longer bar enables the observation of the dynamic propagation regime.

‡: Consistent with loading rates adopted in AFM tests [e.g. Adams et al., 2008].

## 3. Results of simulations

We numerically integrate the equations of motion of the different blocks coupled with the rate and displacement model of the polymeric interface using Newmark's integration method and a predictor-corrector scheme. The detailed numerical approach is provided in Appendix A2. Here we describe some of the quantitative predictions of our model for the different characteristics of crack nucleation and propagation in the fibril-polymer system.

### 3.1 Effect of sacrificial bonds and hidden length (SBHL) system

First, we focus on the effect of sacrificial bonds and hidden length system on the energy dissipation and crack nucleation properties. In each group of simulation, the number of sacrificial bonds (N) per polymer molecule ranges from 0 to 8. The lengths of hidden loops ($L_n$) are chosen from a uniform random distribution.

### 3.1.1 Critical crack size

Figure 6(a) shows the time history for crack propagation along the interface for different number of sacrificial bonds per polymer (N). We define the crack tip position by the location of the farthermost block whose all polymers have been completely detached. The crack length is thus defined as the distance between the edge block and the crack tip.

Under the displacement controlled loading condition considered here, the crack goes through a stable growth phase [Nalla et al., 2004] until it reaches a critical length after which kinetic energy is no longer negligible. This is shown in Figure 6(a). In the stable crack growth, the crack expands slowly at nearly constant speed represented by the initial linear regime in the crack space time plots. As the crack approaches a critical size its propagation speed increases and eventually it diverges signaling that the quasistatic solution is no longer valid and the crack propagation is dynamic. We define the critical crack size as the crack length at the end of the initial linear regime in Figure 6(a). For the simulations shown here, the existence of sacrificial bonds slightly increases the time to dynamic instability (from 0.1063s to 0.1075s) and increases the critical crack size (from ≈ 510 nm to 600 nm). Thus the existence of sacrificial bonds increases the flaw tolerance [Gao et al., 2003] of the system.

In Figure 6(b) we plot the stable crack growth speed as a function of number of sacrificial bonds. The stable crack growth is given by the slope of the initial linear regime in Figure 6(a). The presence of sacrificial bonds reduces the stable crack speed by approximately 1 µm/s (~4.4% of the speed at zero sacrificial bonds). This effect is weakly dependent on the precise number of sacrificial bonds and the crack stable growth speed converges to, on average, 17.5 µm/s for $N \geq 4$.

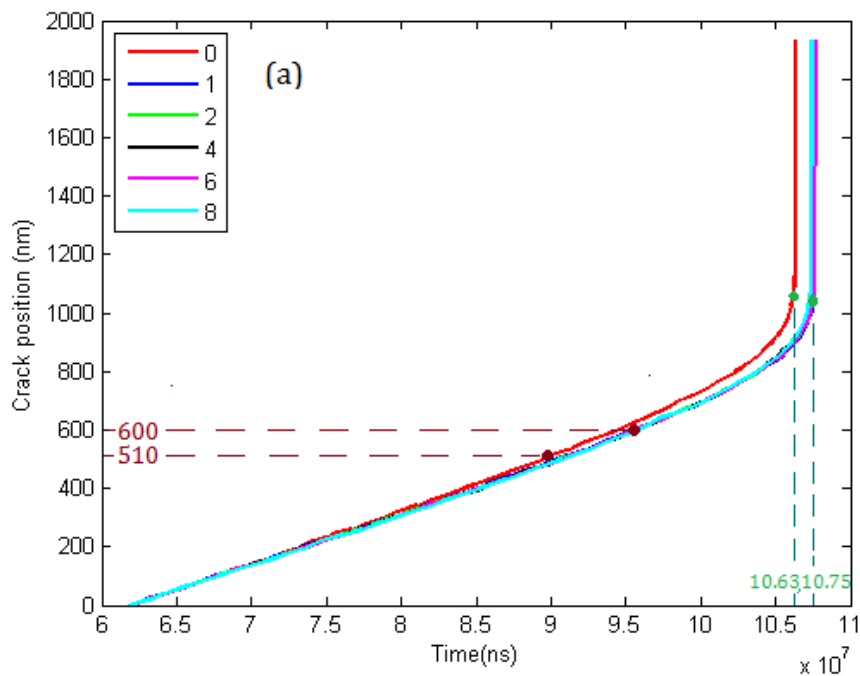

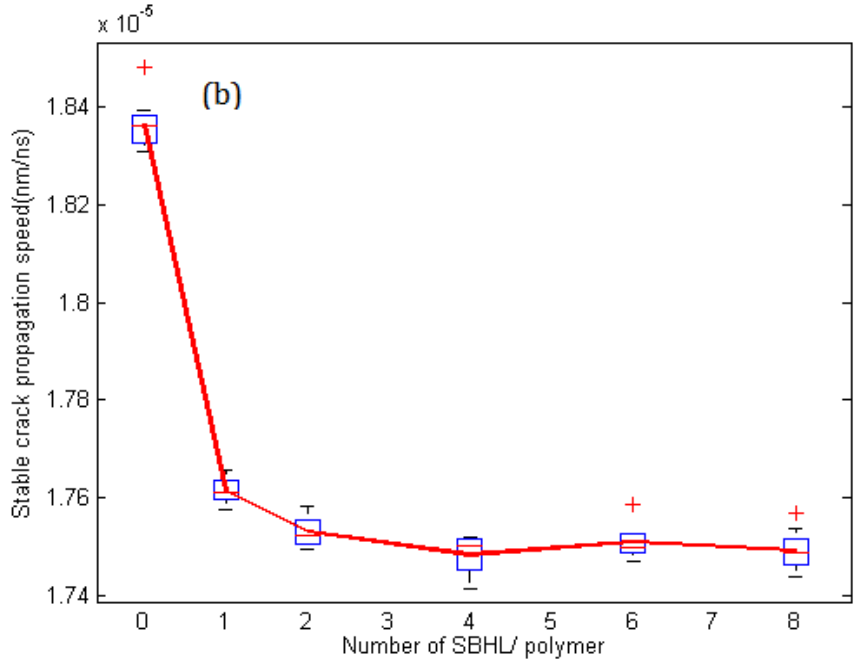

Figure 6: Effect of sacrificial bonds and hidden length system on crack propagation. (a) Crack length (nm) as a function of time (ns) for different number of sacrificial bonds per polymer ranging from 0 to 8. Fully mineralized collagen fibril and polymer density = 25 polymers/nm are used in this model. Increasing the number of sacrificial bonds increases the time to the onset of dynamic instability (signaled by divergent crack propagation speed) and increases the critical crack size (taken as the size corresponding to the end of the stable crack growth at constant speed regime) (b) Average stable crack propagation speed (nm/ns) as a function of number of sacrificial bonds per polymer. The results are averaged over 8 runs. The vertical bars and the red curve indicate one standard deviation and average value, respectively.

### 3.1.2 Energy dissipation

Another indicator of the system response is its fracture toughness. This is measured by how much energy is dissipated as the crack grows and propagates. The higher this energy is, the tougher the system becomes and the more resistant to cracking it is [Park et al., 2009]. There are two types of toughness: initiation toughness and propagation toughness [Nalla, 2004]. The initiation toughness is related to the energy necessary to start the crack growth. The propagation toughness is measured by the energy dissipated during the crack growth. The evolution of energy dissipation as a function of crack position for different numbers of sacrificial bonds per polymer is shown in Figure 7(a). Here, the dissipated energy is calculated as the sum of inelastic work done by all the blocks up to the current time $t$. The initiation toughness increases with increasing the number of sacrificial bonds. The presence of sacrificial bonds also increases the total energy dissipation during the stable crack growth. In Figure 7(b) we compute the total energy dissipation as a function of the number of sacrificial bonds. The energy value is normalized by energy dissipation when there are no sacrificial bonds. As the number of sacrificial bonds increases, the total energy dissipation increases as well. For N=8, the relative increase in energy dissipation is approximately 8.5%.

However, similar to what was shown in Elbanna and Carlson [2013] and Lieou et al., [2013] for individual polymer systems, energy dissipation saturates in the limit of large number of sacrificial bonds. For the simulations shown here, increasing number of sacrificial bonds beyond N=4 has a limited effect on further energy dissipation.

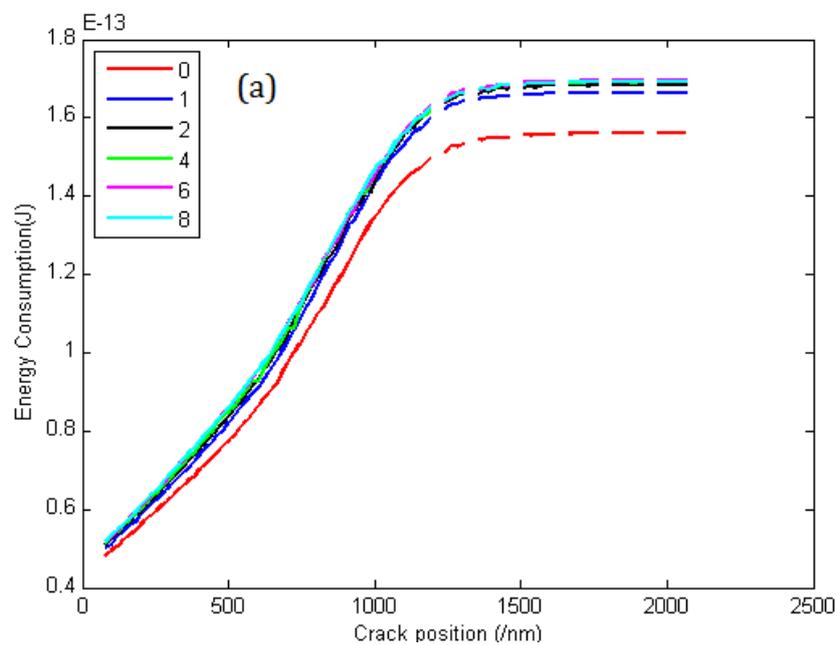

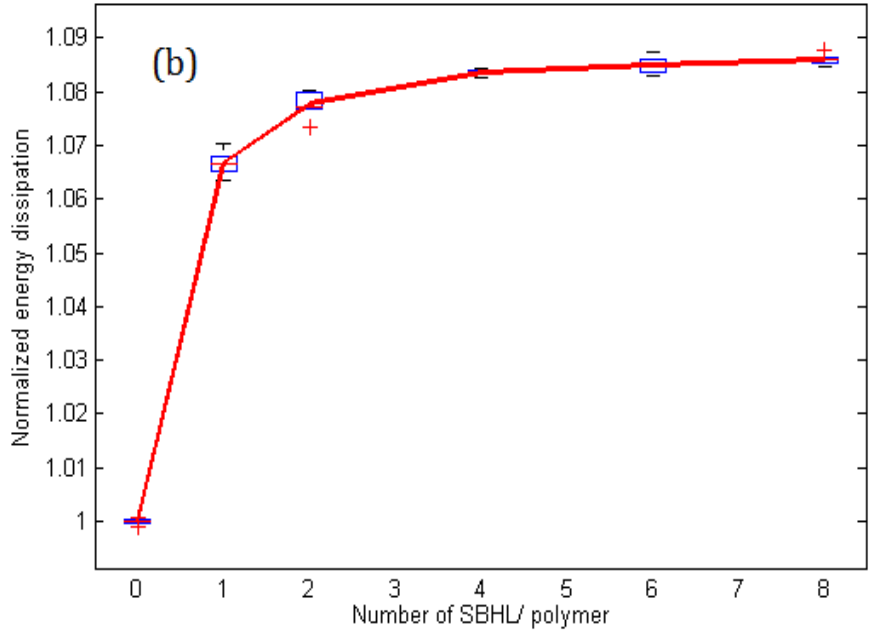

Figure 7: Effect of sacrificial bonds and hidden length system on energy dissipation. (a) Energy consumption (J) versus crack position (nm) for different number of sacrificial bonds per polymer. Fully mineralized collagen fibril and 15 polymers/nm are used in this model. The dotted parts reflect that the mode of crack propagation has become dynamic and energy estimates based on quasistatic analysis are no longer applicable. Increasing the number of bonds increases both the initiation and propagation toughness (b) Normalized total energy dissipation during the entire simulation as a function of number of sacrificial bonds. The results are averaged over 8 runs. The vertical bars and the red curve indicate one standard deviation and average value, respectively.

## 3.2 Effect of polymer density and fibril mineralization

Another important system parameter is polymer density [Elbanna and Carlson, 2013] and degree of mineralization [Jaeger and Fratzl, 2000; Beuhler, 2007; Nair et al., 2013] of the collagen fibril. These are expected to influence the fracture toughness and affect the mechanical features of crack nucleation and propagation.

### 3.2.1 Effect of polymer density on constitutive relation of cohesive interface and energy dissipation:

Developing cohesive law formulations for the polymeric interface is an essential ingredient for multiscale modeling of fracture propagation as it enables the inclusion of small scale physics into macroscopic models [Burr et al., 1988; Vashishth et al., 1997; Ural and Vshishth, 2006, 2007,2007; Yang et al.,2006; Ural, 2009; Ural and Mischinski, 2013] . Figure 8(a) shows the average force, $f_{avg}$, along the interface as a function of the fibril edge displacement for different values of polymer density (5 to 25 polymers/nm). Both the peak force and the edge displacement at failure increase as the polymer density increases. Our results indicate a linear dependence of the peak $f_{avg}$ (denoted by red circles) on the polymer density. The edge displacement at failure is also found to weakly increase as the polymer density increases. This value depends on the length distribution of polymers more than the polymer density itself. However, increasing the number of polymer increases the probability of finding longer polymers leading to larger maximum elongation.

The plots in Figure 8(a) show the general features characteristic of cohesive laws [Park and Paulino, 2001, and references therein]. In the slip strengthening regime, the interfacial force increases as a function of slip up to a critical slip value. Beyond this value, the detachment of polymers lead to progressive softening and slip weakening behavior; the interfacial force decreases with increasing slip.

Variations in polymer density affect fracture toughness. Figure 8(b) shows the time evolution of energy dissipation during crack propagation for different values of polymer density. As the polymer density increases, this leads to (i) an increase in the total value of energy dissipated, and (ii) an increase in the energy dissipation rate as a function of time. This increased energy dissipation leads to a longer slip strengthening region; the displacement at which the peak $f_{avg}$ is achieved increases as the polymer density increases.

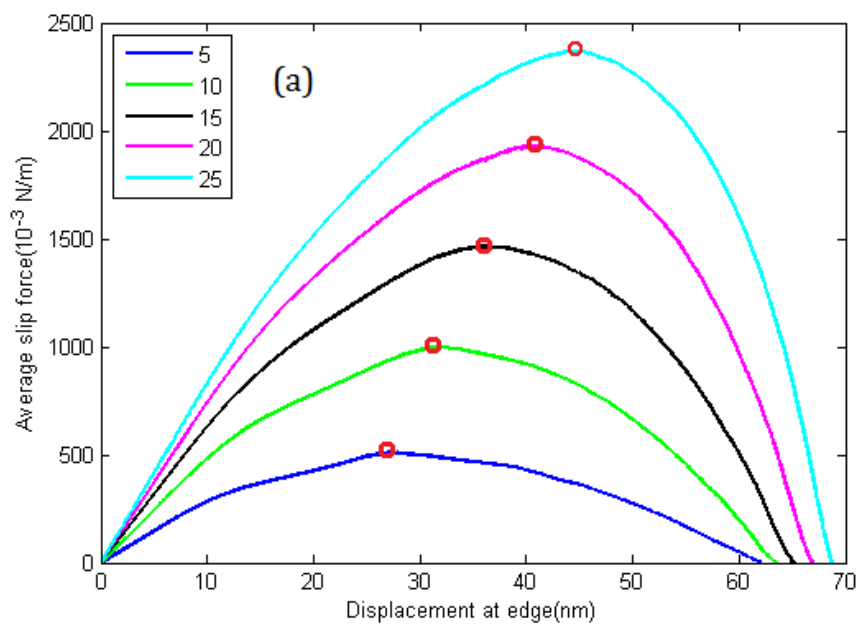

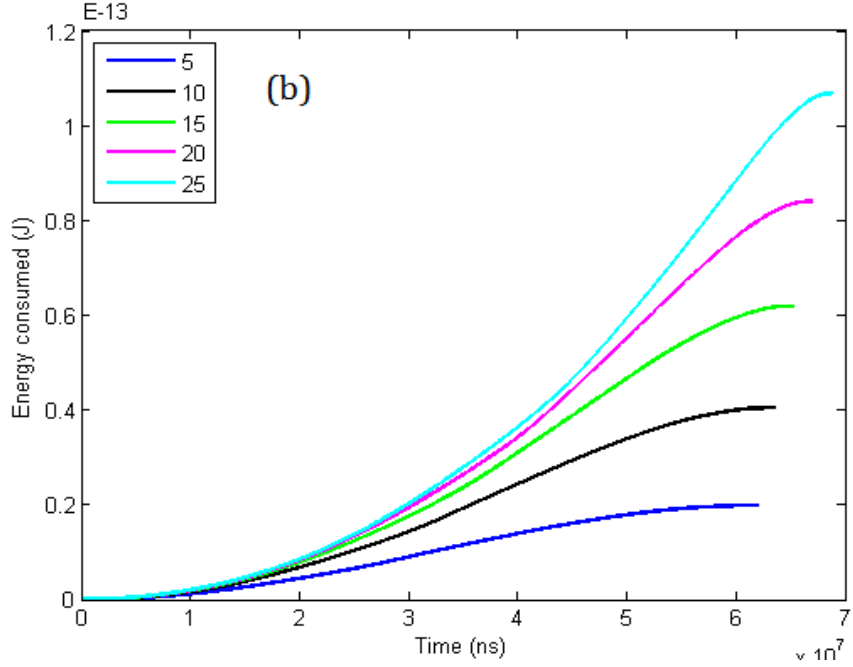

Figure 8: Effect of polymer density on cohesive force and energy dissipation. (a) Average slip force ($10^{-3}$ N/m) as a function of pulling edge displacement (nm), for different polymer densities, from 5 to 25 polymers/nm. The average slip force is the total force at the interface of shear over the length of the interface. Peak $f_{avg}$ are marked with red circles. (b) Energy consumption (J) by polymer system as function of time (ns) for different polymer density ranging from 5 to 25 per nm.

### 3.2.2 Crack propagation

Variations in polymer density are also expected to influence characteristics of crack nucleation and growth, Figure 9(a) shows the crack position as a function of time for different values of polymer density. Increasing the polymer density leads to an increase in the time to crack initiation. For example, for 5 polymers/nm the crack starts to grow at $t = 6.1$ ns, whereas when 25 polymers/nm are used the crack starts to grow at $t = 6.5$ ns. Moreover, polymer density affects the crack growth pattern. At the lowest polymer density, the crack propagates dynamically through the system as soon as it starts. This is reflected by the nearly vertical crack evolution in the space time plot shown in Figure 9(a). On the other hand, at the highest polymer density there is a stable crack growth regime followed by dynamic instability. That is, increasing polymer density leads to an increase in the critical crack size and delays the onset of dynamic crack growth.

These features are further explored in Figure 9(b) where the average crack propagation speed is plotted for various polymer densities. The crack moves approximately 20% to 30% slower for each 5/nm increment of polymer density. Notice that the large variability in case of 5 polymers per nm is reflective of the randomness of the polymer system properties and the significance of discrete effects in the limit of low polymer density.

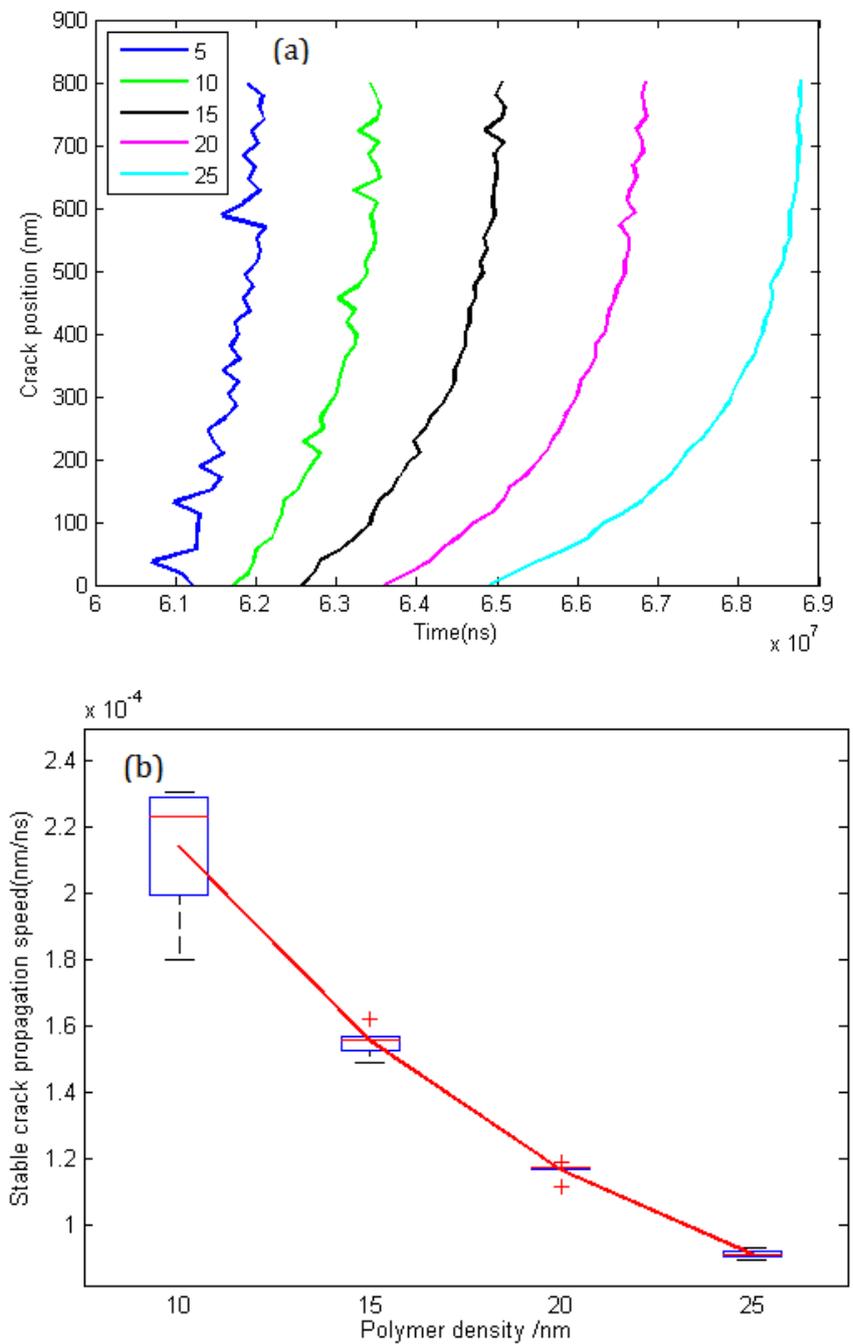

Figure 9: Effect of polymer density on crack propagation (a) Crack length (nm) as a function of time (ns) for different polymer density ranging from 5 to 25/nm. (b) Average crack speed (nm/ns) as function of polymer density (/nm), averaged over 8 runs. The vertical bars and the red curve indicate one standard deviation and average value, respectively.

Fibril mineralization also influences crack growth patterns. Figure 10(a) shows the crack position as a function of time for two limiting cases: an unmineralized case and a fully mineralized case [Beuhler, 2007; Nair et al., 2013]. The mineralization ratio in the fully mineralized case corresponds to what is reported in Beuhler [2007]. The mineralized case appears to be more brittle with shorter rise time to dynamic instability and a slightly smaller critical crack size. In this simulation, the polymeric interface was chosen to be weak enough so that the system fails by sliding along the interface and not by fracture through the fibril. In particular, both the mineralized and the unmineralized fibrils are within their elastic regimes throughout the simulation. In that sense, the difference between the mineralized and the unmineralized cases is that the former is stiffer than the latter. The crack propagation speed depends on different factors such as the interface fracture toughness, the collagen fibril density and the fibril density. Higher fracture toughness leads to lower propagation speeds (see Figure 9(b)) while higher rigidity enables faster crack propagation [Freund, 1990]. Hence, the crack propagates faster as the degree of mineralization increases. This is shown in Figure 10(b) which depicts the average crack speed as a function of the mineralization percentage. The average crack speed rises by 23% from unmineralized to fully mineralized case.

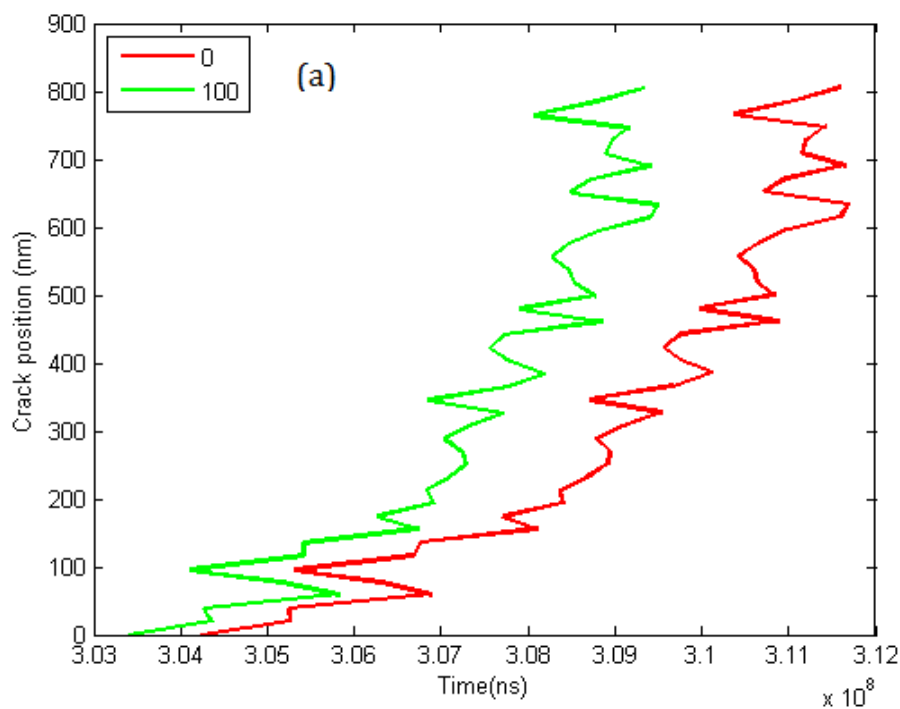

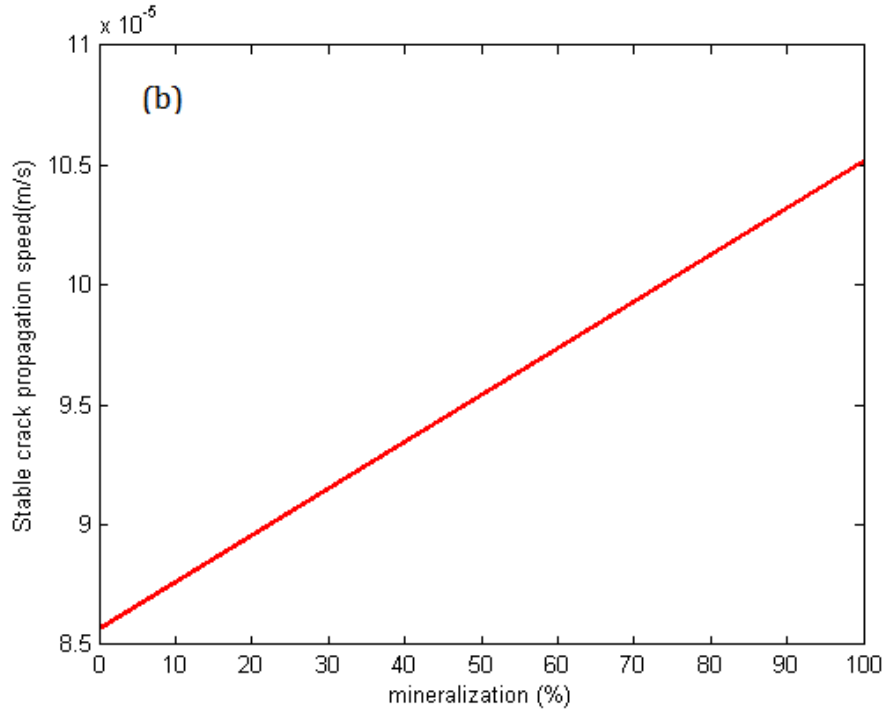

Figure 10: Effect of fibril mineralization on crack propagation. (a) Crack position in length (nm) as a function of time (ns) for different fibril mineralization percentage from 0 to 100. (b) Average crack speed (nm/ns) as function of fibril mineralization (%).

### 3.2.3 Collagen fibril breakage analysis

The results of the previous section are valid when the polymeric interface is weak enough such that the force in the collagen fibril does not exceed its yield strength. In the case of unmineralized collagen fibril, the stiffness and the yielding stress is 33% less than that of the fully mineralized case. If the polymer density is high enough (e.g. 25 polymers/nm), the limit state of tensile failure governs for the case of unmineralized fibril. In this limit, the kinetic energy associated with the fibril softening and breakage is not negligible. To capture these dynamic effects we turn on the dynamic solver a few time steps before the force in the fibril reaches its yield value. The initial conditions for the full dynamic simulation are taken from the results of the last step of the quasistatic analysis.

Figure 11 shows the displacement of the different blocks, representing the collagen fibril, as a function of time. Shortly after the yield force is attained in the first spring, the deformation becomes increasingly localized in the leading edge of the fibril. The collagen fibril eventually breaks in this region before the crack propagates along the interface.

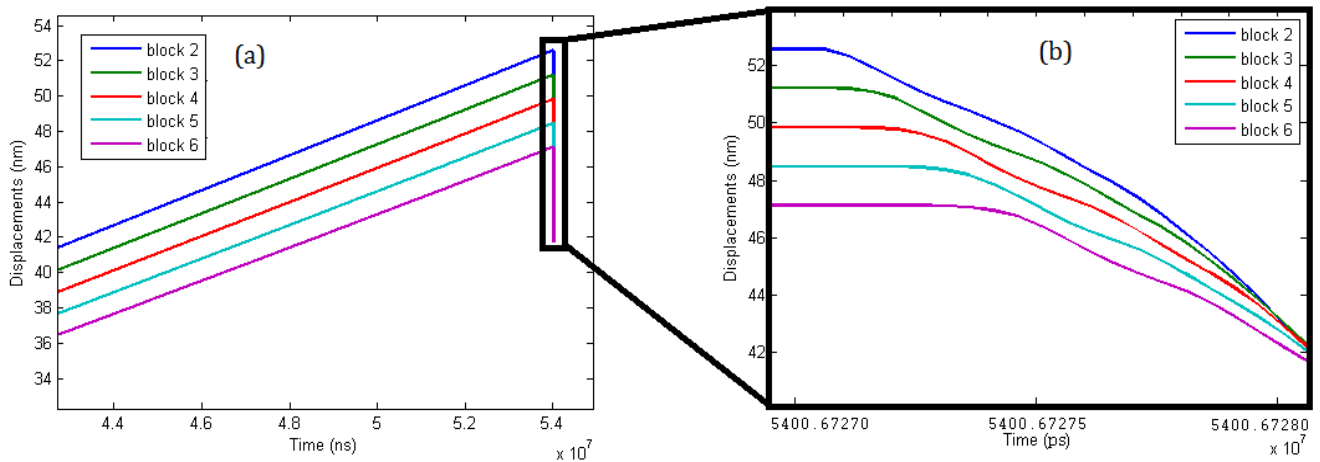

Figure 11: Failure of the system by collagen fibril breakage. (a) Displacement of discretized collagen fibril and (b) zoomed in details at the instant of failure (nm) as a function of time after failure of collagen fibril. Displacements are measured from the initial positions of blocks. Only the first 5 block motions are shown. The displacement of leading block where external force is applied is not shown. Mineralization percentage=0% (pure collagen fibril) and polymer density=25/nm.

Figure 12(a) and 12(b) shows the force displacement relation for the first five springs in our model. Only the first spring was able to reach the yield force and explore the post yielding regime. All other springs remain linearly elastic. The deformation increases in the leading edge of the fibril while the remaining springs unload. As a result, the first spring continues to stretch until it completely fails. The force-extension curve for this spring (Figure 12(a)) is consistent with the idealized constitutive model adopted in Figure 3(b). Other springs unload as their blocks relax to their equilibrium position. This is further shown in Figure 12(c) where the force displacement curves of the polymers attached to the first few blocks are plotted. As the blocks unload, the polymers relax and follow the unloading path shown schematically in Figure 5. Since none of the blocks are detaching, the crack never had a chance to propagate along the interface. This is an example of failure by rupture through the collagen fibril rather than by sliding along the polymeric interface. Unmineralized fibrils are more susceptible to this rupture mode than the mineralized ones.

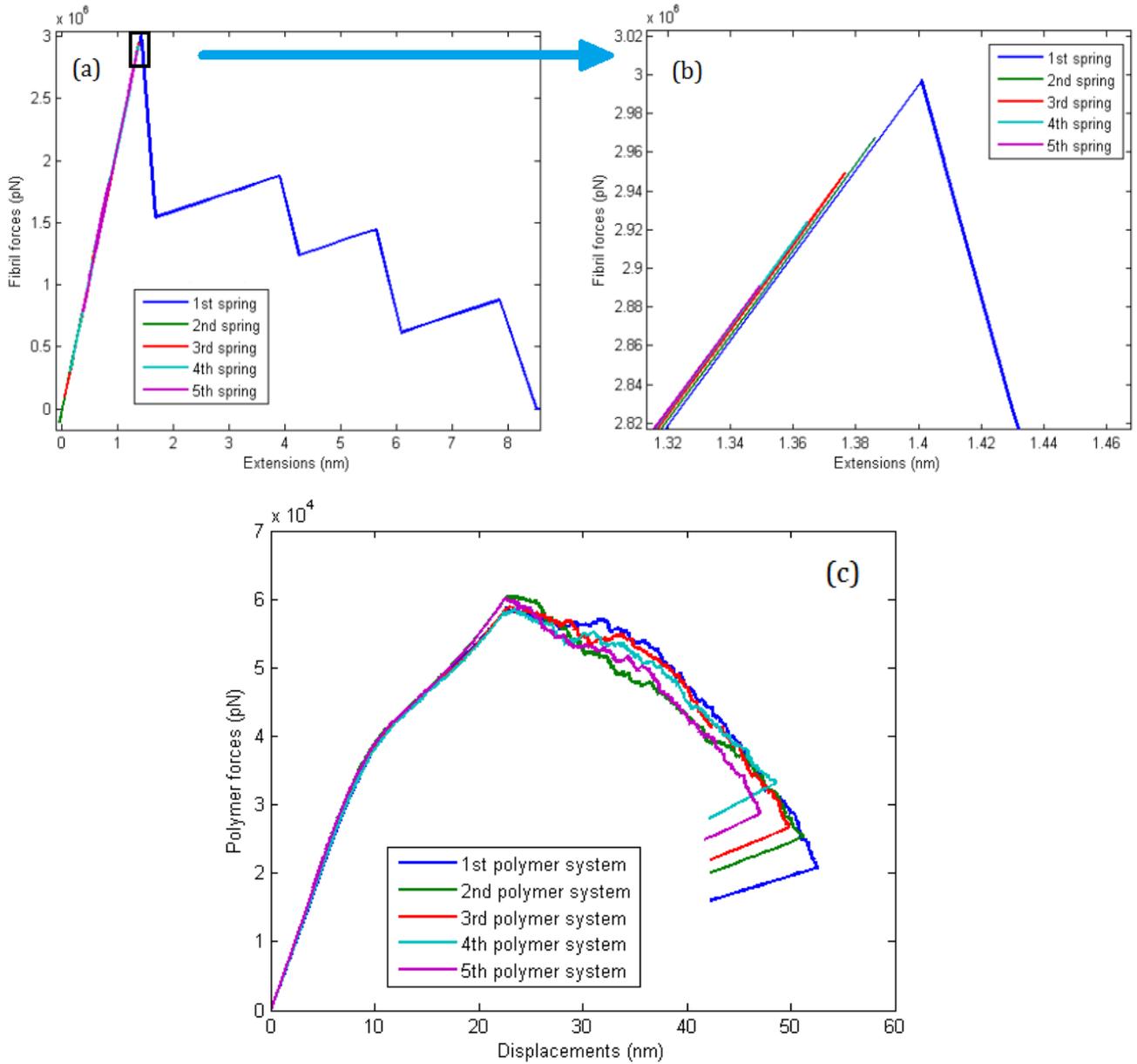

Figure 12: Forces in collagen fibril and polymer system for a case in which the collagen fibril breaks. (a) Forces in idealized collagen fibril springs and (b) zoomed in details at instance of failure (pN) as a function of extension right after failure of collagen fibril. Only forces of the 5 front most fibrils are shown here. The linear part of first spring is not shown. Mineralization percentage=0% and polymer density=25/nm. (c) Forces of polymer system in each discretized block (pN) as a function of extensions (nm) after failure of fibril. Only the forces of the 5 front most polymer systems are shown here. The extensions of polymer system are calculated as the displacements of blocks.

## 4. Discussions

Problems involving dynamics of cohesively held interfaces arise broadly in biological [Ural and Mischinski, 2013 and references therein], engineering [Barenblatt, 1962; Tvergaard and Hutchinson, 1992; Xu and Needleman, 1993; Xia et al., 2006], and geophysical [Scholz, 2002; Lapusta, 2009;

and references therein] applications. Common to all of these applications are fundamental physical processes involving deformation, rupture nucleation, propagation and arrest. In strongly nonlinear problems, like dynamic fracture, small scale instabilities can lead to large scale system fragilities and it is imperative to understand how the microscopic processes influence the crack macroscopic response [Rice, 1983; Carlson and Langer, 1989; Kolmogrov, 1991]. In the case of bone, the internal interfaces between the mineralized collagen fibrils may fail under different loading conditions, and the details of the resulting dynamic rupture can determine whether only a part or the whole of the body part (e.g. knee) will fracture.

In this paper, we focused primarily on interfibrillar sliding as one of the major failure modes in bone structure at the micrometer scale. Other failure modes definitely exist depending on the fibril orientation, bone type and loading conditions. These other modes include for example: delamination (i.e. mode I cracks), twisting (i.e. mode III cracks) and mixed fracture (under combined shear and normal loading of the polymeric interface). The methodology described in this paper is extensible to these other modes and it is expected that the quantitative predictions for the critical crack size, fracture energy and rupture speed may be different. The dominant rupture mode will depend on both the properties of the fibril-polymer system as well as the orientation of the applied loading. For example, delamination may be prevalent in trabecular bone whereas interfibrillar sliding may be dominant in cortical bone.

The separation of the mineralized collagen fibrils under shear or tension is resisted by a special type of polymeric glue that is composed of polymers with the sacrificial bonds and hidden length (SBHL) system [Thompson et al., 2001]. The constitutive response of this glue controls the strength and ductility of the fibril system.

In this paper we used the rate and displacement model developed by Lieou et al. [2013] to model the polymeric interface. We assume that fibril sliding is resisted by a series of quasi-one-dimensional polymer bundles. The idealized SBHL system adapted here represents globular domains within the polymer chains as well as crosslinking between the polymer molecules within these bundles. Crosslinking between the bundles has been neglected here. It is possible; however, that crosslinking may be dense enough to lead to the formation of two-dimensional polymer networks. That will change the force distribution within the polymeric system from the idealized parallel bundles model [see for example Fantner et al., 2006]. In this case, a description based on theory of statistical mechanics that has been previously applied to amorphous materials and thin film lubricants may be more appropriate [Bouchbinder and Langer, 2009]. These theories implement internal state variables to discuss the mechanical evolution of the system. In the case of polymers with SBHL system, the primary state variable will be the number of active bonds. However, as long as the quasi-one-dimensional bundle picture is applicable [e.g. see Fig. 1a] the idealized model presented here, with appropriate choice of distribution of hidden length and strength of sacrificial bonds, is adequate for reproducing the observations of AFM experiments on polymer blobs [e.g. Lieou et al., 2013]. Further investigations are required to explore the influence of extended 2D polymeric structure. Our numerical results show that the sacrificial bonds and hidden length system generally increases energy dissipation and resists crack propagation. The presence of SBHL system increases the system toughness (~ 8.5%), increases the critical crack size that has to be reached before dynamic instability is triggered (~10%), and it also reduces the stable crack growth speed (~5%). The exact numbers depend on the underlying assumption. In particular, we have assumed that the length of the polymers and the hidden loops are drawn from a uniform random distribution. We expect that other probabilistic distributions may lead to different results quantitatively. We think, nonetheless, that the qualitative nature of our findings will continue to hold.

We have also shown that the increase in the polymer density leads to an increase in energy dissipation, peak resistance force and ductility. Smaller polymer density reduces both the initiation and propagation toughness leading to increased brittleness. This is signaled by faster crack propagation as well as smaller critical crack size. Since the number of polymers produced by the osteocytes may decrease as the individual ages [Hansma et al; 2005], this investigation reveals that a possible mechanism for bone toughness degradation with age, other than loss of bone density, is the reduction in polymer density.

Many cohesive law formulations exist in mechanics literature and some of them have been used in the context of multiscale modeling of bone fracture [e.g. Ural and Mischinski, 2013; and references therein]. Our approach is different in the sense that the cohesive law is derived based on physical principles. The constitutive law parameters (such as peak force, maximum elongation and fracture energy) vary in response to variations in the internal variables (e.g. polymer density and number of sacrificial bonds) in a self-consistent way. Hence, the proposed approach has a more predictive power and is capable of integrating small scale physics in multiscale simulations without prior assumptions on the specific shape of the cohesive law.

Similar to other systems [e.g. Gao et al., 2003], we have shown that there exists a critical crack size beyond which crack propagation becomes dynamic. Determining this critical crack size is important for both medical and engineering applications. In particular, it may be used for assessing fracture risk in bone by determining how close the current crack size is to the critical one. Moreover, the critical crack size sets an important length scale to be considered in biomimetic material design. For example, to increase fracture resistance, polymeric interfaces should be continuous only for distances smaller than the critical crack length. In that aspect, staggered or random distribution of the fibrils may be preferred to the more regular brick and mortar geometry [Jaeger and Fratzel, 2000].

The degree of mineralization of the collagen fibril is another factor controlling the fracture properties of bone. We have found that within the elastic regime of the fibril, the average crack propagation speed along the interface increases as the percentage of mineralization increases. On the other hand, for unmineralized fibrils and high polymer density the system fails by strain localization within the fibril rather than by slip along the interface. This mode of failure is brittle and prevents the full utilization of the SBHL system. With aging, the degree of mineralization is reduced [Nalla et al., 2004; Koester, 2011; Nair et al., 2013] and this may be another mechanism for frequent bone fractures in the elderly.

In this paper, we adopted the approximation that mineral plates fill the gaps between the tropocollagen molecules. The exact distribution of mineralization is a subject of debate. The distribution, however, may have strong impact on the mechanical response of the fibril, especially its compressive strength. Here, we assume that the fibril buckles once it reaches the yield stress in compression and consequently loses its compressive strength completely. Better understanding of the mineralization distribution may allow us to track the fibril behavior in the post-buckling regime.

The model we developed in this paper is one-dimensional. We reduced the problem to its basic ingredients related to fibril elasticity and polymer toughness. We were able to develop some constraints on the basic fracture response of the collagen fibrils at the microscale. Three-dimensional effects for bone are important, however, and the geometrical complexity is an essential ingredient for toughness [Ritchie et al., 2009]. This current study represents the first step towards building these more complicated models.

The model proposed in this paper provides predictions for critical crack size, stable crack propagation speed and energy dissipation for a fibril-polymer system under different conditions. These predictions are derived using a mathematically consistent procedure (integrating Newton's second law) that implements constitutive models that were validated independently by different experiments. For example, the rate and displacement model of Lieou et al. (2013) reproduces many of the AFM experimental observations on glue molecules [e.g. Admas et al., 2008; and references therein] including logarithmic rate dependence of strength, time dependent healing and irregular force drops. On the other hand, the mechanical model of the fibril [Beuhler, 2007] was validated using Synchrotron diffraction studies of Gupta et al. [2004]. To the best of our knowledge there are yet no experiments that have been done to explore the fracture behavior of a single fibril-polymeric interface system at the microscale with which we can compare our predictions to. While the different parts of the model have been validated independently, we do believe there is a need for developing experiments that can probe the fracture response at that scale. In this respect, we believe that extending novel techniques such as the scratch test [Akono et al., 2011] to the microscale may provide valuable insight into the fracture toughness of bone at the scale of collagen fibril. Mechanical testing at the microscale alongside with Scanning Electron Microscope Imaging can provide information about crack development and speed. This may be done by loading the fibril in incremental steps, stopping the experiment after each step, and imaging the deformation patterns. Eventually, a multiscale framework of fracture that integrates the model proposed here as its building block, will link the micro and macroscales and provide more opportunities for validation through classical fracture mechanics techniques [e.g. Ritchie et al. 2009] and medical diagnostics.

Further extension of this study includes investigating arrays of collagen fibrils in two and three dimensions using the finite element method including other failure modes such as delamination, twisting and mixed mode fractures. This will enable the investigation of the characteristics of the wave field generated by the crack propagation and the influence of array geometry on crack propagation, crack arrest and energy dissipation for bone structure. It will also help us better understand the fundamental mechanics of deformation in bone which will eventually help in developing better biomimetic materials.

## Acknowledgments

The authors are grateful to two anonymous reviewers for their thoughtful and constructive comments which greatly enhanced the manuscript. The authors are also grateful to Jean Carlson, Paul Hansma, Charles Lieou and Darin Peetz for constructive discussions on SBHL systems. This research is supported by the Department of Civil and Environmental Engineering (CEE Innovation grant) at the University of Illinois (Urbana-Champaign).

# Appendix

This appendix reviews the displacement and rate constitutive model for polymers with SBHL systems [Lieou et al., 2013] that is implemented in this paper. We also describe the details of the numerical scheme used in the integration of the equations of motion.

**A1. Rate dependence breakage mechanism**

To account for pulling rate effects, Bell's theory is applied [Bell, 1978; Lieou et al., 2013]. Assuming a double well potential for the bond, the rates of bond formation and breakage depend on the applied force. In particular, the rates of these two events have exponential dependence on the pulling force and transition state distance. That is, for sacrificial bonds:

$$k_f = \alpha_0 \exp\left(\frac{F\Delta x_f}{k_B T}\right) \tag{A1.1}$$

$$k_b = \beta_0 \exp\left(-\frac{F\Delta x_b}{k_B T}\right) \tag{A1.2}$$

For end bonds, we obtain:

$$k_f^{end} = \alpha_e \exp\left(\frac{F\Delta x_f^{end}}{k_B T}\right) \tag{A1.3}$$

$$k_b^{end} = \beta_e \exp\left(-\frac{F\Delta x_b^{end}}{k_B T}\right) \tag{A1.4}$$

Here, $F = F(x)$ is the polymer force given by Eq. (1), $\Delta x_f$ and $\Delta x_b$ are the distances to the transition state; $\alpha_0$ and $\beta_0$ are, respectively, inverse time scales which describe the rate at which bond breakage and formation events occur at zero pulling force. The same applies for end bonds. The master equation for the change of the number of bonds is given as:

$$\frac{dN_b^*}{dt} = -k_f N_b + k_b N_f \tag{A1.5}$$

where $N_b^*$ is the continuous version of integer $N_b$, representing the number of sacrificial bonds at a given instant of time. $N_f = N - 2N_b$ is the number of free sites to form potential sacrificial bonds, with $N = L/b$ being the number of sites. The same applies for end bonds as:

$$\frac{dN_e^*}{dt} = -k_f^{end} N_e + k_b^{end}(1 - N_e) \tag{A1.6}$$

The condition for bond breakage is thus satisfied if:

$$\int_{t_1}^{t_2} \left(k_f(F(x))N_b - k_b(F(x))N_f\right)dt = 1 \tag{A1.7}$$

$$\int_0^{t_c} k_f^{end}(F(x))dt = 1 \tag{A1.8}$$

For the parameters here, we followed Lieou et al. [2013] and choose the following in our simulations: $\alpha_0 = 0.3$ s$^{-1}$, $\beta_0 = 0.003$ s$^{-1}$, $\alpha_e = 0.1$ s$^{-1}$, b = 0.1 nm, $\Delta x_f = 0.25$ nm, $\Delta x_b = 0.1$ nm, and $\Delta x_e = 0.15$ nm.

**A2. Implementation of dynamic and quasistatic analysis**

The equations of motion of individual blocks take the following form:

$$m_i \ddot{x}_i + F_{cf_{i+1,i}}(x_{i+1} - x_i) + F_{cf_{i,i-1}}(x_i - x_{i-1}) + F_{p_i}(x_i, \dot{x}_i) = 0 \tag{A2.1}$$

Here, $F_{cf_{i+1,i}}$ and $F_{cf_{i,i-1}}$ are the forces on the $i^{th}$ block due to its motion relative to the $i+1$ and $i-1$ blocks, respectively. These forces are a function of the relative displacement between the adjacent blocks and their values are determined from Figure 3(b). $F_{p_i}(x_i, \dot{x}_i)$ is the polymer force acting on the $i^{th}$ block. It is a function of the block absolute displacement and velocity.

We numerically integrate the system in time using Newmark-beta method [Newmark, 1959]:

$$\dot{x}_{t+\Delta_t} = \dot{x}_t + \left[(1-\alpha)\ddot{x}_t + \alpha \ddot{x}_{t+\Delta_t}\right]\Delta_t. \tag{A2.2}$$

$$x_{t+\Delta_t} = x_t + \dot{x}_t \Delta_t + \left[\left(\frac{1}{2} - \beta\right)\ddot{x}_t + \beta \ddot{x}_{t+\Delta_t}\right]\Delta_t^2. \tag{A2.3}$$

In the above equations, $t$ is the current time step, $t + \Delta_t$ is the next time step, $\Delta_t$ is the time step size, and coefficients $\alpha$ and $\beta$ are set as 1/2 and 1/4 respectively. Since the acceleration at the next time step is not known, we implement a predictor-corrector scheme to solve Eqns. (A2.2) and (A2.3). All the above equations together with Eqn. (1) represent a highly nonlinear system with strong coupling between forces, displacements and velocities. At time $t$, the displacement and polymer forces for all the blocks are known. We use this information to compute the instantaneous accelerations of all the blocks at time $t$. To estimate the response of the system at time $t + \Delta_t$ we initially assume that the acceleration of each block is constant during the interval $[t, t + \Delta_t]$. Using Eqns. (A2.2) and (A2.3) a predicted value for the displacement and velocity of each block at time $t + \Delta_t$ is determined. Using the predicted values we estimate the polymer and collagen forces acting on each block. We use these values to compute the new acceleration magnitude for each block at time $t + \Delta_t$. This latter value is not, in general, equal to the constant acceleration assumed previously. We then use Newmark-beta method (Eqns. (A2.2) and (A2.3)) to compute corrected values for displacements and velocities. We repeat the process until the errors between the predicted and computed displacements (velocities) are sufficiently small ($10^{-6}$ of the latest predicted value).

We adopt an adaptive time stepping algorithm. To detect bond breakage events, we integrate Eqns. (A1.7) and (A1.8) numerically using a trapezoidal rule. We also compute the changes in the polymer force due to the breakage of a sacrificial or end bond. If the value of integration of either Eqn. (A1.7) or (A1.8) is greater than $1 + \varepsilon$, where $\varepsilon$ is a prescribed small number, there is a probability that the drop in polymer force may not be accurately represented. In this case, the time step is reduced to half its original value and the calculation is repeated. Only if the drop in the polymer force is much smaller than the total force in the polymer system do we tolerate integration outcomes exceeding 1 by values slightly larger than $\varepsilon$. In this case, the effect of the discontinuity is negligible in the force displacement curve. This enables us to use larger time steps without compromising the accuracy of the constitutive law.

Two approaches are implemented in this paper: a quasistatic approach to model stable crack growth, and a fully dynamic approach to model fracture of the collagen fibril. In the former, the inertia term in Eqn. (A2.1) is set to zero. The time dependent loading (imposed displacement rate at the end) as well as the rate dependence of the polymer force are still included. In this case, we solve for static equilibrium, using Eqn. (A2.1) but with the left hand side set to zero, at each time step. In the dynamic approach, the inertia term is included and integrated numerically. This is appropriate for tracking dynamic crack propagation and strain localization in the fibril during the post-yielding stage. At the instant of yielding, we switch to the dynamic solver and repeat the last few simulation steps from the quasistatic analysis. The initial conditions for the dynamic analysis are taken from the solution corresponding to the last step in the quasistatic solution. Both the dynamic and quasistatic analysis yield identical results for the stable crack growth regime as long as the collagen fibril remain elastic. The comparison of two schemes is shown in Figure A2.1.

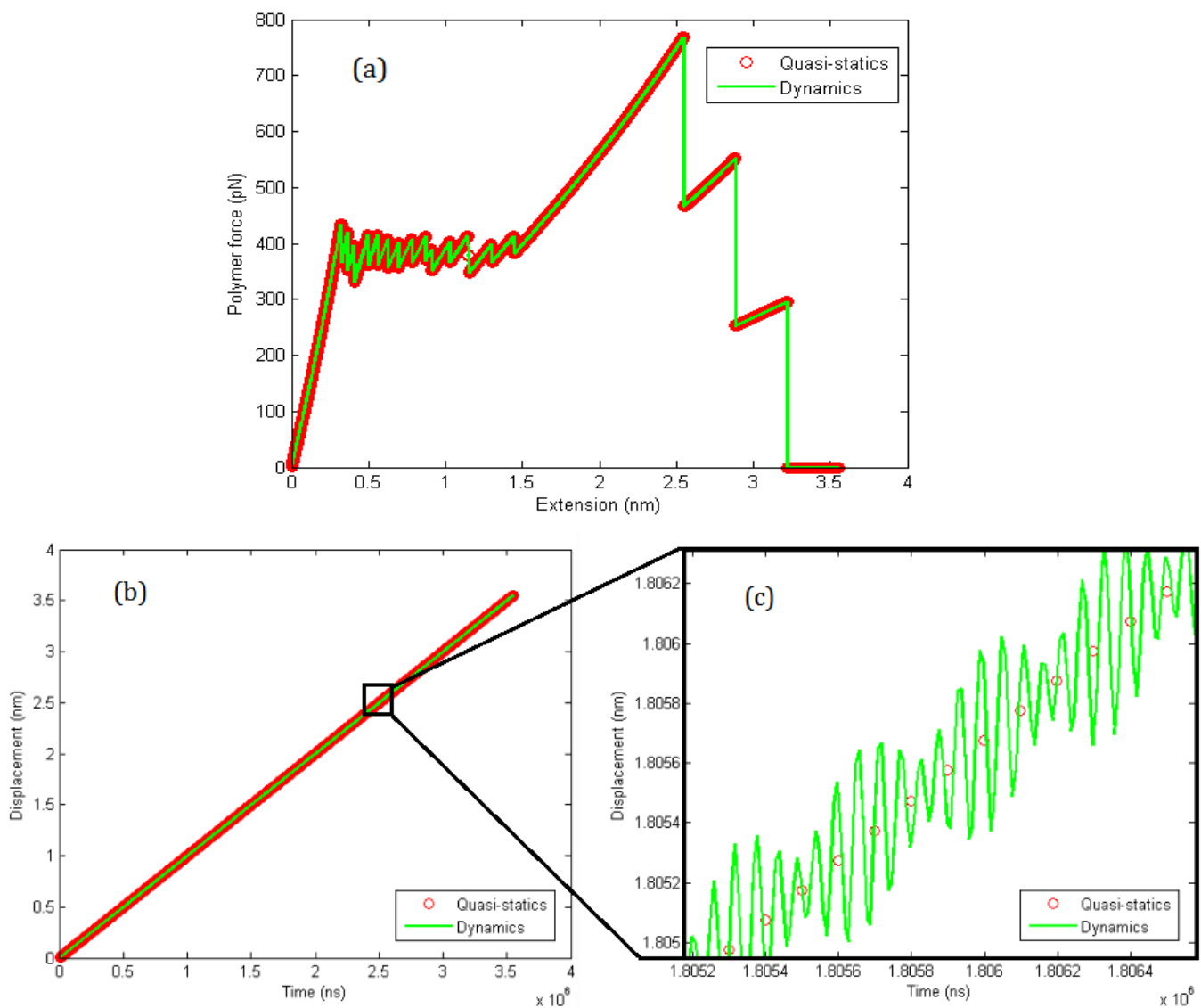

Figure A2.1: Comparison of quasistatic analysis and dynamic analysis on a benchmark problem. (a) Polymer forces (pN) as a function of extension (nm). Both scheme yield identical results. (b) Motions (nm) of fibril block as a function of time (ns) and (c) zoomed-in details of motions. The motion of quasistatic analysis can be idealized as the averaged dynamical motion.